\documentclass[prx,aps,amsmath,amssymb,floatfix,twocolumn,longbibliography]{revtex4-2}

\usepackage{bm}
\usepackage{color}
\usepackage{xfrac}
\usepackage{braket}
\usepackage{hyperref}
\usepackage{placeins}
\usepackage{graphicx}
\usepackage[utf8]{inputenc}
\usepackage{amsmath,amsfonts,amssymb}

\def\eqq#1{Eq.~(\ref{#1})}

\def\eq#1{(\ref{#1})}

\def\ff#1{Fig.~\ref{#1}}

\def\s#1{Section~\ref{#1}}

\def\c#1{~\cite{#1}}

\def\cc#1{Ref.~\cite{#1}}

\def\av#1{\left \langle #1 \right \rangle}

\def\beq{\begin{equation}}
\def\eeq{\end{equation}}
\def\bea{\begin{eqnarray}}
\def\eea{\end{eqnarray}}

\def\tf{t_{\rm f}}
\def\f{_{\rm f}}
\def\i{_{\rm i}}

\definecolor{blue}{rgb}{0,0,1}

\makeatletter
\def\maketitle{
\@author@finish
\title@column\titleblock@produce
\suppressfloats[t]}
\makeatother

\begin{document}

\title{Learning protocols for the fast and efficient control of active matter}
\author{Corneel Casert}
\email{ccasert@lbl.gov}
\affiliation{Molecular Foundry, Lawrence Berkeley National Laboratory, 1 Cyclotron Road, Berkeley, California 94720, USA}
\affiliation{Department of Physics and Astronomy, Ghent University, 9000 Ghent, Belgium}
\author{Stephen Whitelam}
\email{swhitelam@lbl.gov}
\affiliation{Molecular Foundry, Lawrence Berkeley National Laboratory, 1 Cyclotron Road, Berkeley, California 94720, USA}

\begin{abstract}

We show that it is possible to learn protocols that effect fast and efficient state-to-state transformations in simulation models of active particles. By encoding the protocol in the form of a neural network we use evolutionary methods to identify protocols that take active particles from one steady state to another, as quickly as possible or with as little energy expended as possible. Our results show that protocols identified by a flexible neural-network ansatz, which allows the optimization of multiple control parameters and the emergence of sharp features, are more efficient than protocols derived recently by constrained analytical methods. Our learning scheme is straightforward to use in experiment, suggesting a way of designing protocols for the efficient manipulation of active matter in the laboratory.

\end{abstract}
\maketitle
\section{Introduction}

Active particles extract energy from their surroundings to produce directed motion~\cite{ramaswamy2010mechanics,romanczuk2012active,marchetti2013hydrodynamics,bechinger2016active}.
Natural active particles include groups of animals and assemblies of cells and bacteria~\cite{cavagna2014bird,elgeti2015physics,needleman2017active}; synthetic active particles include active colloids and Janus particles~\cite{zottl2016emergent,buttinoni2012active}. Active matter, collections of active particles, displays emergent behavior that includes motility-induced phase separation~\cite{Cates_2015,10.1039/9781839169465-00107}, flocking~\cite{toner2005hydrodynamics,chate2020dry}, swarming~\cite{be2019statistical}, pattern formation~\cite{liebchen2017collective,liebchen2018synthetic}, and the formation of living crystals~\cite{palacci2013living}.

Recent work has focused on controlling such behavior by creating active engines~\cite{PhysRevE.102.010101, fodorActiveEnginesThermodynamics2021,kumariStochasticHeatEngine2020,cocconi2023optimal,saha2023information,holubec2020active,holubec2020underdamped,datta2022second,gronchi2021optimization,pietzonka2019autonomous}, controllably clogging and unclogging microchannels~\cite{caprini2020activity}, doing drug delivery in a targeted way~\cite{ghosh2020active, luo2018micro}, and creating  microrobotic swarms with controllable collective behavior~\cite{rubenstein2014kilobot,yigit2019programmable,balda2022playing}. For such applications, efficient time-dependent protocols are important~\cite{guptaEfficientControlProtocols2023,guery-odelinDrivingRapidlyRemaining2023,norton2020optimal,chennakesavalu2021probing,shankar2022optimal}. Methods for identifying efficient protocols, such as reinforcement learning, have been used to optimize the navigation of active particles in complex environments~\cite{monderkamp2022active,nasiri2022reinforcement,nasiri2023optimal} and induce transport in self-propelled disks using a controllable spotlight~\cite{falk2021learning}.

For purely diffusive (passive) molecular systems, analytic methods allow the identification of optimal time-dependent protocols for a range of model systems~\cite{schmiedl2007optimal, gomez2008optimal,blaber2021steps,zhong2022limited}. These results establish that rapidly-varying and discontinuous features are common components of optimal protocols, and are useful for benchmarking numerical approaches~\cite{whitelamDemonMachineLearning2023,engelOptimalControlNonequilibrium2023}. However, active-matter systems are more complicated to treat analytically than passive systems, requiring the imposition of protocol constraints in order to make optimization calculations feasible for even the simplest model systems. Two recent papers derive control protocols for confined active overdamped particles by assuming that protocols are slowly varying and smooth~\cite{davisActiveMatterControl2023} or have a specific functional form~\cite{baldovinControlActiveBrownian2023}. In this paper we show numerically that relaxing these assumptions leads to more efficient control protocols for those systems. The protocols we identify can be rapidly varying and are not smooth, showing jump discontinuities similar to those seen in overdamped passive systems.

To learn protocols to control active matter we use the neuroevolutionary method described in Refs.~\cite{whitelam2020learning,whitelam2021neuroevolutionary, whitelamDemonMachineLearning2023, whitelamHowTrainYour2023}, which we adapted from the computer science literature~\cite{GA, mitchell1998introduction, such2017deep}. Briefly, we encode a system's time-dependent protocol~\footnote{It is also straightforward within this scheme to consider a feedback-control protocol, by considering a neural network ${\bm g}_{\bm \theta}(t/\tf,{\bm v})$, where ${\bm v}$ is a vector of state-dependent information\c{whitelamDemonMachineLearning2023}} in the form ${\bm g}_{\bm \theta}(t/\tf)$. Here ${\bm g}$ is the output vector of a deep neural network, corresponding to the control parameters of the system (which in this paper consist of the activity of the particles and the spring constant of their confining potential), ${\bm \theta}$ is the set of neural-network weights, $t$ is the elapsed time of the protocol, and $\tf$ is the total protocol time. We apply the protocol to the system in question, and compute an order parameter $\phi$ that is minimized when it achieves our desired objective (such as inducing a state-to-state transformation while emitting as little heat as possible). The neural-network weights ${\bm \theta}$ are then iteratively adjusted by a genetic algorithm in order to identify the protocol whose associated value of $\phi$ is as small as possible.

This approach is a form of deep learning -- in the limit of small mutations and a genetic population of size 2 it is equivalent to noisy gradient descent on the objective $\phi$~\cite{whitelam2021correspondence} -- and so comes with the benefits and drawbacks of deep learning generally. Neural networks are very expressive, and if trained well can identify ``good'' solutions to a problem, but these solutions are not guaranteed to be optimal\c{hornik1989multilayer,bahri2020statistical}. We must therefore be pragmatic, and (as with other forms of sampling) verify that protocols obtained from different starting conditions and from independent runs of the learning algorithm are consistent. Consequently, we call the protocols identified by the algorithm ``learned'' rather than ``optimal''.  In general, we have found the method to be easy to apply and to solve the problems we have set it: we have benchmarked the method -- see Refs.~\cite{whitelamDemonMachineLearning2023, whitelamHowTrainYour2023} and Fig.~\ref{fig:seifert} in the Supplementary Information (SI) -- against exact solutions~\cite{schmiedl2007optimal}  and other numerical methods~\cite{engelOptimalControlNonequilibrium2023,rotskoff2015optimal,gingrich2016near,zhong2022limited}. In this paper we use it to produce protocols that are closer to optimal than the protocols obtained by other methods~\cite{davisActiveMatterControl2023, baldovinControlActiveBrownian2023}. 

The neuroevolutionary learning algorithm uses information that is experimentally accessible. And so while in this paper we have learned protocols for the control of simulation models (these protocols could then be applied to experiment if the simulation model is a good enough representation of the experiment), the learning algorithm can also be applied directly to experiment. The following results therefore indicate the potential of this method for learning control protocols for active systems generally.

\section{Active particle in a trap of variable stiffness}\label{sec:heat}

\begin{figure*}[]
\includegraphics[width=\textwidth]{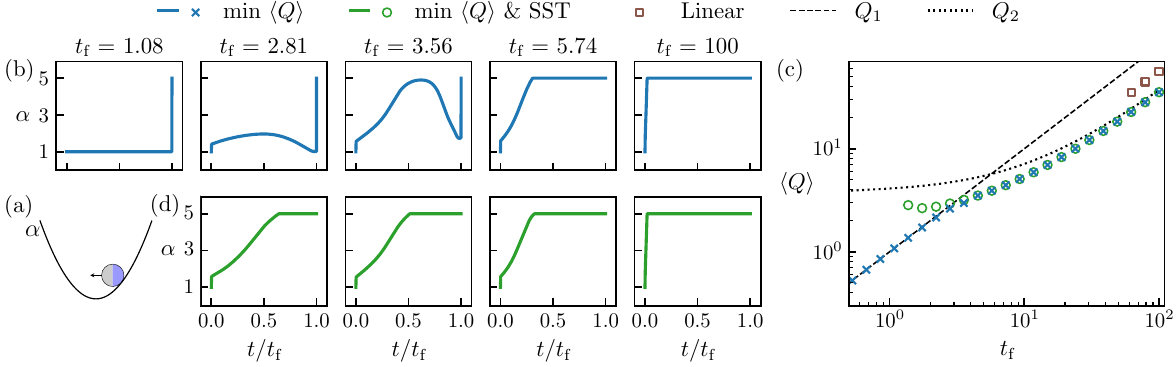} 
\caption{Protocols $\alpha(t)$ controlling the stiffness of a trap confining an active Ornstein-Uhlenbeck particle. (a) Schematic of the model. (b) Neural-network protocols $\alpha(t)$ that minimize $\langle Q \rangle $, \eqq{heat}, for different protocol lengths $\tf$. (c) Heat associated with the protocols. The dotted and dashed lines correspond to the values $Q_1$, \eqq{q1}, and $Q_2$, \eqq{q2}, respectively. (d) As (b), but requiring minimum heat while enacting a state-to-state transformation (SST); see \eqq{phi2}. Model parameters: $D=1, \mu = 1, \tau = 1, D_1 = 2$.}
\label{fig1}
\end{figure*}

In this section we consider the problem of Section IIIA of Ref.~\cite{davisActiveMatterControl2023}, a single active Ornstein-Uhlenbeck particle~\cite{PhysRevE.103.032607, irreversibility2022, bonilla2019active} in a one-dimensional harmonic trap of stiffness $\alpha(t)$. A schematic of this model is shown in \ff{fig1}(a). The particle has position $r$ and self-propulsion velocity $v$. It experiences overdamped Brownian motion with diffusion constant $D$ and mobility $\mu$, such that
\beq
\dot{r}(t) = v(t) -\mu\alpha\, r(t) + \sqrt{2D}\eta(t).
\eeq
Here $\eta$ is a Gaussian white noise term with zero mean and unit variance. The self-propulsion velocity $v$ follows an Ornstein-Uhlenbeck process with persistence time $\tau$ and amplitude $D_1$, such that $\av{v} = 0$ and $\av{v(t)v(t')} = D_1 \tau^{-1} e^{-|t-t'|/\tau}$. The parameter $D_1$ is zero in the passive limit. 

The trajectory-averaged heat associated with varying $\alpha(t)$ from $\alpha\i$ to $\alpha\f$ in time $\tf$ is~\cite{davisActiveMatterControl2023, irreversibility2022, fodorActiveEnginesThermodynamics2021}
\begin{align}
    \begin{split}
\label{heat}
\av{Q}=&\frac{1}{2} \left( \alpha\i x\i-\alpha\f x\f \right)+ \frac{1}{2} \int_0^{\tf} {\rm d} t\, \dot{\alpha}(t) x(t)  \\
&+\frac{D_1 \tf}{\tau \mu}-\int_0^{\tf} {\rm d} t\, \alpha(t) y(t).
\end{split}
\end{align}
Here $\av{\cdot}$ denotes an average over dynamical trajectories, and we have defined $x \equiv \av{r^2}$ and $y \equiv \av{ r v}$.  For time-dependent quantities $q(t)$ we use the notation $q\i \equiv q(0)$ and $q\f\equiv q(\tf)$ to denote initial and final values.
The first line of \eqq{heat} is the passive heat (minus the change in energy plus the work done by changing the trap stiffness), and the second line is the active contribution to the heat. The first term on the second line is constant for fixed $\tf$ (describing the heat dissipated to sustain the self-propelled motion), and plays no role in selecting the protocol. 

For a given protocol $\alpha(t)$, the time evolution of $x$ and $y$ is given by the equations\c{davisActiveMatterControl2023}
\begin{align}
    \begin{split}
\label{evo}
\frac{1}{2} \dot{x}(t)+ \mu \alpha(t) x(t) &= y(t)+D, \,\,{\rm and} \\
\tau \dot{y}(t)+\gamma(t) y(t)&=D_1,
\end{split}
\end{align}

where $\gamma(t) \equiv 1+ \mu \tau \alpha(t)$. The system starts in the steady state associated with the trap stiffness $\alpha\i$, and so its initial coordinates are
\beq
\label{ss}
x\i = \frac{1}{\alpha\i\mu}\left(\frac{D_1}{\gamma\i} + D\right) \, \, {\rm and} \, \, \, y\i = \frac{D_1}{\gamma\i}.
\eeq

Ref.~\cite{davisActiveMatterControl2023} sought protocols that carry out the change of trap stiffness $\alpha\i = 1 \to \alpha\f = 5$ with minimum mean heat, \eqq{heat}. The theoretical framework used in that work assumes that protocols $\alpha(t)$ are smooth and are not rapidly varying (see \s{heat_si} for a discussion of this point). Here we revisit this problem using neuroevolution. We find that heat-minimizing protocols are not in general slowly varying or smooth, but can vary rapidly and can display jump discontinuities. The protocols we identify produce considerably less heat than do the protocols identified in~\cc{davisActiveMatterControl2023} (see Fig.~\ref{fig:heat_SI}).

To learn a protocol $\alpha(t)$ that minimizes heat, we encode a general time-dependent protocol using a deep neural network. We choose the parameterization
\beq
\label{ansatz}
\alpha_{\bm \theta}(t) = \alpha \i + (\alpha \f - \alpha\i) (t/\tf) + g_{\bm \theta}(t/\tf),
\eeq
where $g$ is the output of a neural network whose input is $t/\tf$. We constrain the neural network so that $\alpha \i \leq  \alpha_{\bm \theta}(t) \leq \alpha \f$, meaning that it cannot access values of $\alpha$ outside the range studied in Ref.~\cite{davisActiveMatterControl2023}~\footnote{When we relax this constraint we find protocols that produce less heat, in general, than the protocols that observe the constraint. We impose the constraint to allow us to make contact with~\cc{davisActiveMatterControl2023}, and because experimental systems have constraints on the maximum values of their control parameters}. Initially the weights and output of the neural network are zero, and so we start by assuming a protocol that interpolates linearly with time between the initial and final values of $\alpha$. We train the neural network by genetic algorithm to minimize the order parameter $\phi=\av{Q}$, given by \eqq{heat}, which we calculate for a given protocol by propagating \eq{evo} for time $\tf$, using a forward Euler discretization with step $\Delta t = 10^{-3}$. An example of the learning process is shown in Fig.~\ref{fig:heat_SI}(a).

In \ff{fig1}(b) we show, for the choice $D_1=2$, that heat-minimizing protocols learned by the neural network vary between a step-like jump at the final time, for small values of $\tf$, and a step-like jump at the initial time, for large values of $\tf$. For intermediate values of $\tf$ we observe a range of protocol types. These protocols include non-monotonic and rapidly-varying forms, and show jump discontinuities at initial and final times.

 The heat associated with the final-time step protocol is just that associated with the initial steady state, and is 
\beq
\label{q1}
Q_1 = \frac{D_1 \tf}{\mu\tau(1+ \alpha \i \mu \tau)}.
\eeq 
The heat associated with the initial-time jump protocol can be calculated from Eqs.~\eq{heat} and~\eq{evo}, and is 
\begin{align}
\label{q2}
    \begin{split}
Q_2 =& \frac{\alpha \f}{2} \left(x\i-x_2(\tf)\right) + \frac{D_1 \tf}{\mu \tau(1+\alpha \f \mu \tau)}  \\ 
&-\frac{D_1 \tau \alpha\f}{\gamma\f}\left( \frac{1}{\gamma \i}- \frac{1}{\gamma \f}\right)\left( 1-{\rm e}^{-\gamma\f \tf/\tau}\right),
\end{split}
\end{align}
where
\begin{align}
    \begin{split}
x_2(t)\equiv& (x\i-x\f) {\rm e}^{-2 \mu \alpha \f t}+x_{\rm ss}  \\
&+ 2D_1 \left(\frac{1}{\gamma\i}-\frac{1}{\gamma\f}\right) \left(2 \mu \alpha \f-\frac{\gamma\f}{\tau} \right)^{-1} \\
&\times \left({\rm e}^{-\gamma\f t/\mu}-{\rm e}^{-2 \mu \alpha \f t}\right).
\end{split}
\end{align}
(Note that $x_{\rm ss}$ is given in \eqq{ss2}.) For large $\tf$ we have
\beq
Q_2 \approx \frac{D_1 \tf}{\mu\tau(1+ \alpha \f \mu \tau)},
\eeq 
which is the heat associated with the final steady state.

In \ff{fig1}(c) we show that the heat values associated with the trained neural-network protocols interpolate, as a function of $\tf$, between the values $Q_1$ and $Q_2$. Our conclusion is that this optimization problem is solved by protocols that are rapidly varying, have a variety of functional forms, and display jump discontinuties. As shown in Fig.~\ref{fig:heat_SI}, these protocols produce values of heat considerably smaller than those associated with the protocols derived in~\cite{davisActiveMatterControl2023}. (In that figure we also show that it is possible to construct smooth but rapidly-varying protocols that can produce values of heat arbitrarily close to the discontinuous protocols identified by the learning procedure.)

The protocols just described are valid solutions to the heat-minimization problem defined in~\cite{davisActiveMatterControl2023}. However, some of them are not meaningful in experimental terms. For instance, for small values of $\tf$, the heat-minimizing protocol is a step function at the final time. This protocol is a solution to the stated problem, but effects no change of the system's microscopic coordinates. All the heat associated with the subsequent transformation of the system is ignored, simply because we have stopped the clock.

We therefore argue that it is more meaningful to search for protocols that minimize heat subject to the requirement of a state-to-state transformation. That is, we require that a specified change in the system's state has occurred. We therefore modify the problem studied in~\cc{davisActiveMatterControl2023} to search for protocols that minimize the mean heat \eq{heat} caused by a change of trap stiffness $\alpha\i = 1 \to \alpha\f = 5$, {\em subject to the completion of a state-to-state transformation} (SST) between the initial steady state \eq{ss} and that associated with the final-time value of $\alpha\f$, 
\beq
\label{ss2}
x_{\rm ss} = \frac{1}{\alpha\f\mu}\left(\frac{D_1}{\gamma\f} + D\right) \, \, {\rm and} \, \, \, y_{\rm ss} = \frac{D_1}{\gamma\f}.
\eeq
As before, we impose the experimentally-motivated constraint $\alpha \i \leq  \alpha_{\bm \theta}(t) \leq \alpha \f$. 

To solve this dual-objective problem we choose the evolutionary order parameter
\beq
\label{phi2}
\phi=\Delta +c \, \, {\rm if}\, \Delta \geq \Delta_0 \, \, {\rm and} \, \, \phi = \av{Q} \, {\rm otherwise}. 
\eeq 
Here $\Delta^2 \equiv (x\f-x_{\rm ss})^2+(y\f-y_{\rm ss})^2$ measures the difference between the final-time system coordinates and their values \eq{ss2} in the final steady state; $\Delta_0=10^{-3}$ is the tolerance with which we wish to achieve this steady state~\footnote{Protocols and heat values depend weakly on the value of this threshold, but not in a way that affects our general conclusions.}; and $c=100$ is an arbitrary constant whose only role is to make the first clause of \eq{phi2} always larger than the second. Minimizing \eq{phi2} will minimize heat emission for a protocol $\alpha(t)$ that in time $\tf$ effects a state-to-state transformation within the precision $\Delta_0$. 

In \ff{fig1}(d) we show protocols that minimize heat while achieving SST. These protocols have a variety of of forms, which involve rapidly-varying portions and jump discontinuities, and that tend, for large $\tf$, to the initial-time jump form. For times $\tf \lesssim 1.3$ the learning algorithm could not identify a protocol that could achieve SST. 

The heat emission associated with these protocols is shown in panel (c). The time $\tf$ for which least heat is emitted is about $\tf=1.74$, for this choice of $D_1$. (For heat optimization alone, the minimum heat is $\av{Q}=0$, and is shown by \eqq{q1} to occur at time $\tf=0$, a conclusion different to that drawn in Fig. 3 of \cc{davisActiveMatterControl2023}. This strange result follows from the fact that the instruction to minimize heat comes with no requirement that the system change state.) 

For comparison, we show the heat emission associated with the linear protocol $\alpha_{\rm lin}(t) = \alpha \i + (\alpha \f - \alpha\i) (t/\tf)$ (square symbols). The linear protocol emits considerably more heat than learned protocols (note the log scale of the figure), and fails to achieve SST for times $\tf \lesssim 60$. 

We conclude that the model of the confined active particle studied in~\cite{davisActiveMatterControl2023} is best controlled by protocols $\alpha(t)$ that are in general rapidly varying and exhibit jump discontinuities -- similar to protocols for overdamped passive systems -- whether the goal is to minimize heat or to do so while also inducing SST. We note that while the evolutionary training of the neural network is a numerical procedure, the protocols it identified allowed us to derive analytic results for the minimum heat produced for sufficiently small and large trajectory lengths, \eqq{q1} and \eqq{q2} respectively.

\section{Active particle of variable activity in a trap of variable stiffness}\label{sec2}
\subsection{State-to-state transformation in least time}
\begin{figure*}[t]
\includegraphics[width=\textwidth]{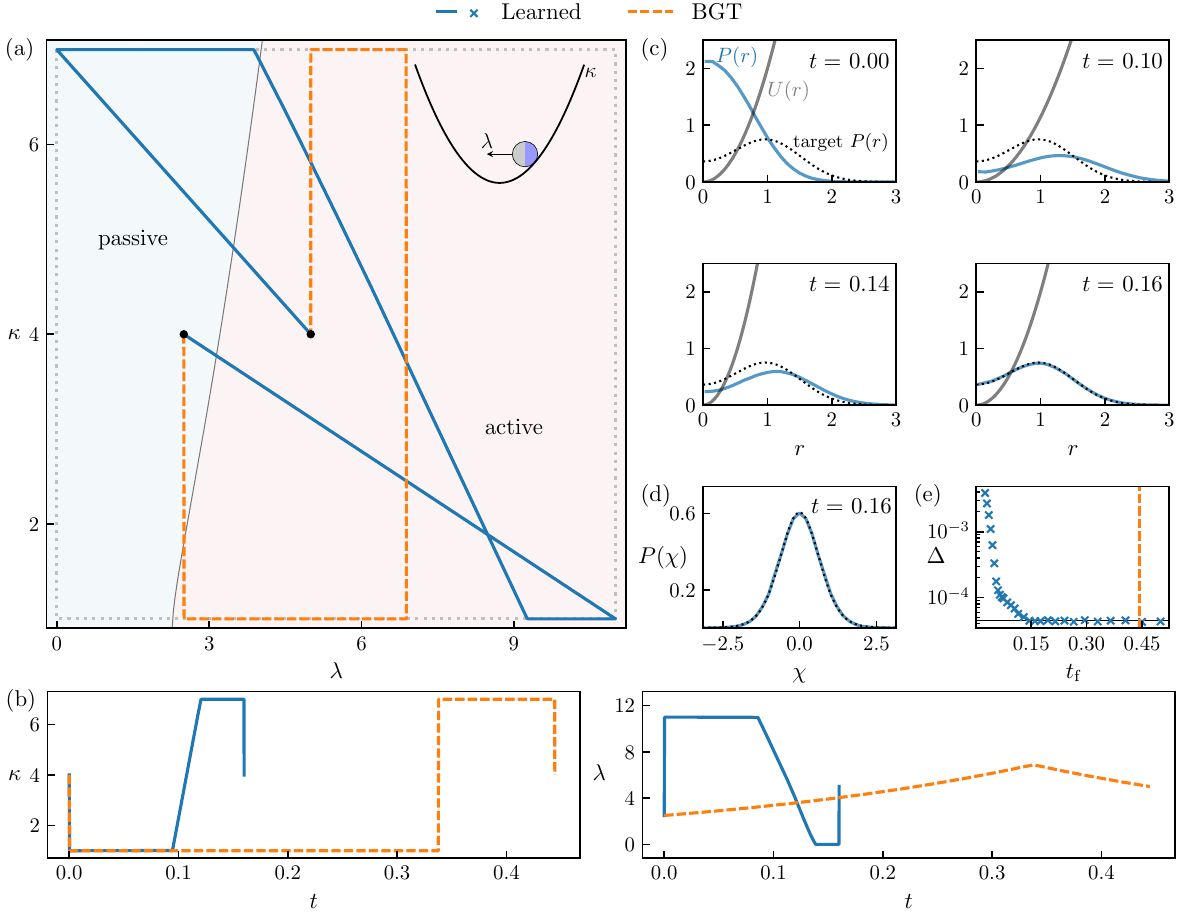} 
\caption{State-to-state transformation for a confined active Brownian particle with control protocol $(\lambda(t),\kappa(t))$. (a) Parameteric protocols from~\cite{baldovinControlActiveBrownian2023} (orange, ``BGT'') and this work (blue). The latter achieves the transformation about three times more rapidly than the former. The black dots denote initial and final points; the dotted line denotes the bounds for the control parameter values. The schematic in the top-right corner is a one-dimensional schematic of this two-dimensional system. (b) The protocols of panel (a) as a function of time $t$. (c) Temporal evolution of the radial distribution function $P(r)$ together with the target distribution (dotted line) and the potential $U(r)$, for the learned protocols shown in (a,b). (d) Final-time distribution for $\chi$. The dotted line is the target distribution. (e) Mean-squared error $\Delta$ between the final distribution and the exact solution, averaged over $10^5$ trajectories. The dashed vertical line is the transformation time for the protocol of~\cite{baldovinControlActiveBrownian2023}, and the horizontal line is the value of $\Delta$ associated with that protocol.}
\label{fig2}
\end{figure*}

In this section we consider the problem of Ref.~\cite{baldovinControlActiveBrownian2023}, an active Brownian particle confined by a two-dimensional harmonic potential $U({\bm \rho})=\frac{1}{2} k {\bm \rho}^2$ with stiffness $k$. The particle is described by the position vector $\bm{\rho} = (\rho \cos\phi, \rho\sin\phi)$ and orientation $\theta$, and moves in the direction $ \hat{\bm{e}}(\theta)=(\cos\theta, \sin\theta)$ with constant speed $u_0$. Its dynamics is described by the Langevin equation
\begin{align}
\label{lang1}
    \begin{split}
    \frac{\mathrm{d}\bm{\rho}}{\mathrm{d}\tau} &= u_0 \hat{\bm{e}}(\theta) - \mu k \bm{\rho} + \sqrt{2D_t}\bm{\xi}_r(\tau)\\ 
 \frac{\mathrm{d}\theta}{\mathrm{d}\tau} & = \sqrt{2D_\theta}\xi_\theta(\tau),
    \end{split}
\end{align}
where $\tau$ is the time; $\mu$ is the mobility; $D_t$ and $D_\theta$ are translational and rotational diffusion coefficients, respectively; and $\bm{\xi}_r(\tau)$ and $\xi_\theta(\tau)$ are Gaussian white noise terms with zero mean and unit variance. Upon introducing the dimensionless variables $\bm{r} \equiv \bm{\rho}\sqrt{D_\theta/D_t}$ and $t \equiv \tau D_\theta$, \eqq{lang1} reads
\begin{align}
\label{lang2}
    \begin{split}
    \frac{\mathrm{d}\bm{r}}{\mathrm{d}t} &= \lambda \hat{\bm{e}}(\theta) - \kappa \bm{r} + \sqrt{2}\bm{\xi}_r(t)\\ 
 \frac{\mathrm{d}\theta}{\mathrm{d}t} & = \sqrt{2}\xi_\theta(t),
    \end{split}
\end{align}
where $\kappa \equiv \mu k/D_\theta$ and $\lambda \equiv u_0/\sqrt{D_\theta D_t}$ are dimensionless versions of the spring constant and the self-propulsion speed ($\lambda$ is the P\'eclet number). These dimensionless variables are the control parameters of the problem.

The steady-state probability distribution function $\mathcal{P}_{\rm ss}(r,\chi)$ of the system depends only on $r\equiv |\bm{r}|$ and $\chi \equiv \theta - \phi$, and is known exactly~\cite{malakarSteadyStateActive2020}. The steady state associated with the control-parameter choices $\kappa$ and $\lambda$ can be classified as passive or active (Fig.~\ref{fig2}(a)): in the passive phase, the radial probability distribution $P(r)$ is peaked at the trap center, while in the active phase it is peaked at $r > 0$.

This model system is motivated by experiments involving spherical Janus particles, whose self-propulsion speed can be tuned through light intensity~\cite{buttinoni2012active}, confined in a trap constructed by acoustic waves~\cite{takatori2016acoustic}. For a typical experimental setup the control parameters are bounded as $0 \le \lambda \le 11$ and $1 \le \kappa \le 7$~\cite{buttinoni2012active,takatori2016acoustic,baldovinControlActiveBrownian2023}. 

The problem described in Ref.~\cite{baldovinControlActiveBrownian2023} is to find a time-dependent protocol $(\lambda(t),\kappa(t))$ that obeys the bounds of the previous paragraph and that minimizes the time $\tf$ required to transform the distribution $\mathcal{P}(r,\chi)$ from a passive steady state at $(\lambda\i,\kappa\i) = (2.5,4)$ to an active one at  $(\lambda\f,\kappa\f) = (5,4)$. Using an ansatz constrained so that the distribution $\mathcal{P}(r(t),\chi(t))$ has at all times the form of the steady-state distribution $\mathcal{P}_{\rm ss}(r,\chi)$ with effective control-parameter values, the authors of that paper found a protocol that completed the state-to-state transformation in time $\tf \approx 0.44$. This protocol is shown in \ff{fig2}(a,b).

With a neural-network ansatz for the protocol $(\lambda(t),\kappa(t))$ we find that the state-to-state transformation can be achieved about three times as rapidly; see \ff{fig2}(a,b). For a simulation of fixed time $\tf$ we use a genetic algorithm to train the neural network to minimize the order parameter $\phi=\Delta$, the mean-squared error between the target distribution $\mathcal{P}^\star_{\rm ss}(r,\chi)$ associated with the control-parameter values $(\lambda\f,\kappa\f)$ and the distribution $\mathcal{P}(r(\tf),\chi(\tf))$ obtained at the end of the simulation. The latter was calculated from $10^5$ independent trajectories of \eq{lang2} under a given neural-network protocol.

The protocol learned by the neural network for time $\tf=0.16$ is shown in \ff{fig2}(a,b), together with the protocol of Ref.~\cite{baldovinControlActiveBrownian2023}. Both show sharp jumps in trap stiffness, decreasing it abruptly to its smallest possible value. The neural network protocol achieves the transformation more quickly because it also enacts a sharp jump in activity, setting it to the maximum possible value (the constraints imposed in Ref.~\cite{baldovinControlActiveBrownian2023} mean that if one control parameter achieves its maximum value in an abrupt way, the other is not free to do so). Near the end of the learned protocol both parameters are abruptly changed to their final values. 

In \ff{fig2}(c) we show the temporal evolution of $P(r)$ for the learned protocol. Starting from an initial distribution peaked at the origin, the peak of $P(r)$ overshoots the peak of the target distribution (they are not at that time of the same shape). The peak of $P(r)$ is later brought back toward the target when stiffness and activity are set to their maximal and minimal values, respectively. Subsequently, both are set to their final values.

In \ff{fig2}(d) we show the final-time distribution of $\chi$ for the learned protocol, which matches the target distribution.

In \ff{fig2}(e) we show the value of $\Delta$ obtained by protocols trained at various fixed simulation times $\tf$. For times $\tf \gtrsim 0.15$, the learned protocol produces a small constant value of $\Delta$ consistent with the value produced by the protocol of Ref.~\cite{baldovinControlActiveBrownian2023} (horizontal line). For times $\tf \lesssim 0.15$ the value of $\Delta$ increases sharply with decreasing $\tf$, indicating that the state-to-state transformation cannot be achieved with the same precision. 

\begin{figure*}[t]
\includegraphics[width=\textwidth]{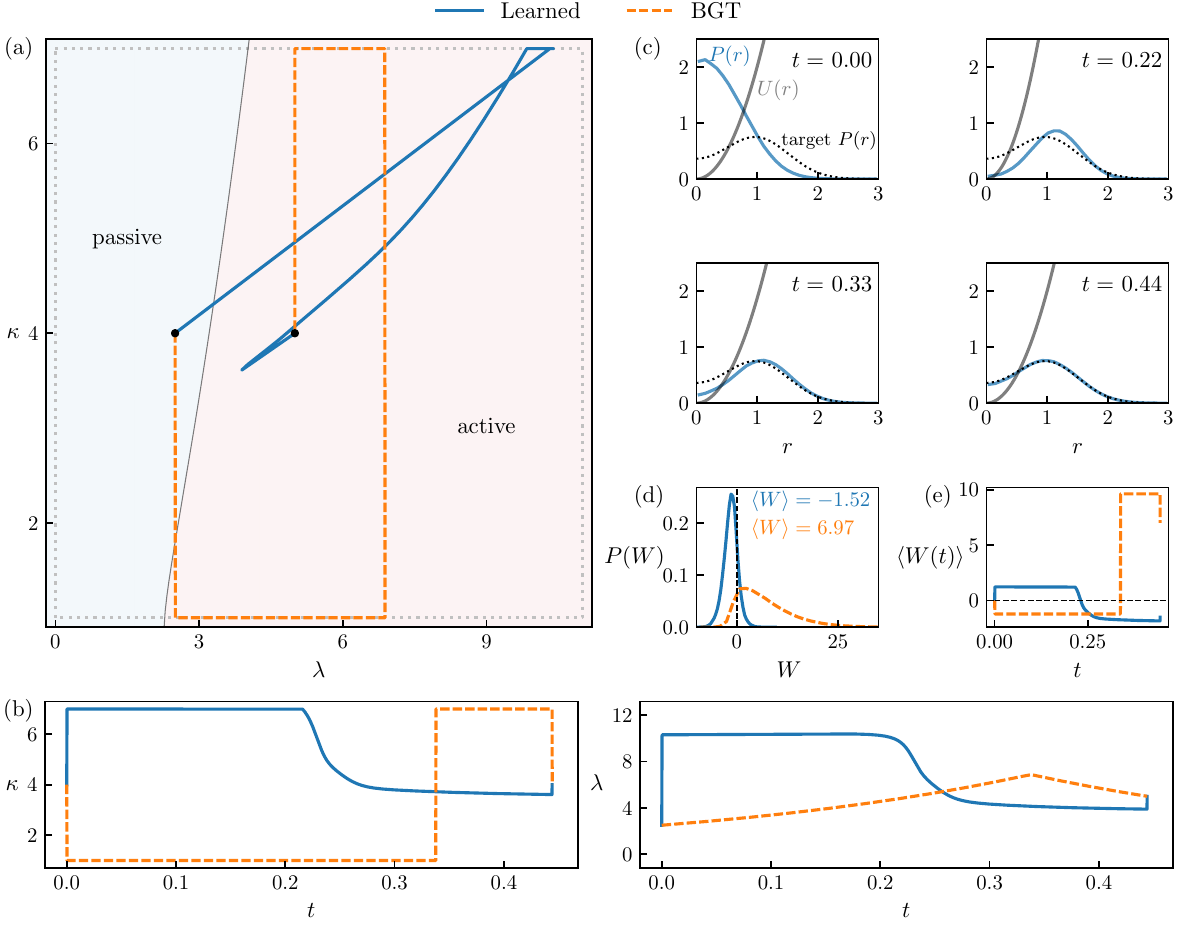} 
\caption{Similar to \ff{fig2}, but now the learning algorithm is told to enact the state-to-state transformation of \ff{fig2}, in time $\tf= 0.44$, while minimizing work done; see \eqq{phi3}. Panels (a--c) are analogous to those of \ff{fig2}. (d) Distribution of work for the two protocols. (e) Mean work as a function of time for the two protocols.}
\label{fig3}
\end{figure*}

\subsection{State-to-state transformation with work extraction}
\label{sec3b} 
It is possible to extract work during the state-to-state transformation. Setting $\tf=0.44$, the transformation time of the protocol of Ref.~\cite{baldovinControlActiveBrownian2023}, we used a genetic algorithm to train a neural network to minimize the objective
\beq
\label{phi3}
\phi=\Delta +c \, \, {\rm if}\, \Delta \geq \Delta_0 \, \, {\rm and} \, \, \phi = \av{W} \, {\rm otherwise}. 
\eeq 
Here $\Delta_0$ is the mean-squared error associated with the protocol of Ref.~\cite{baldovinControlActiveBrownian2023} (calculated using $10^5$ trajectories), and $c=100$ is an arbitrary constant whose only role is to make the first clause of \eq{phi3} always larger than the second. The quantity $\av{W}$ is the mean work, in units of $\mu/D_t$, given by
\beq
    \langle W\rangle =  \int_0^{t_{\rm{f}}} \mathrm{d}t\, \dot{\kappa} \left< \frac{\partial U}{\partial \kappa}\right>= \frac{1}{2}  \int_0^{t_{\rm{f}}} \mathrm{d}t\, \dot{\kappa} \left< r^2\right>.
    \label{eq:work}
\eeq
Minimizing \eq{phi3} will minimize the mean work associated with a protocol $(\lambda(t),\kappa(t))$ that in time $\tf$ effects the state-to-state transformation to a precision $\Delta_0$. 

The protocol learned in this way is shown in \ff{fig3}(a,b), together with the protocol of Ref.~\cite{baldovinControlActiveBrownian2023}. Panels (c) show the effect of the learned protocol on the radial probability distribution. The neural-network protocol increases $\kappa$ to its maximum value at the beginning of the protocol. Doing so costs work, but only small amounts because the system is initially in a passive phase and so $\langle r^2\rangle$ is small. The protocol also increases $\lambda$ to a large (but sub-maximal) value, which begins to drive the distribution into the active phase, so increasing $\langle r^2\rangle$. Subsequently, $\kappa$ is decreased to its target value, causing a decrease of energy and allowing net extraction of work.

\ff{fig3}(d) shows the work distributions $P(W)$ associated with the learned protocol and that of Ref.~\cite{baldovinControlActiveBrownian2023}. The latter results in a broad distribution of work values, and on average requires a large input of work to enact the transformation. By contrast, the work distribution obtained using the learned protocol is sharply peaked at a negative value, and the mean work is negative. 
\begin{figure*}[t]
\includegraphics[width=\textwidth]{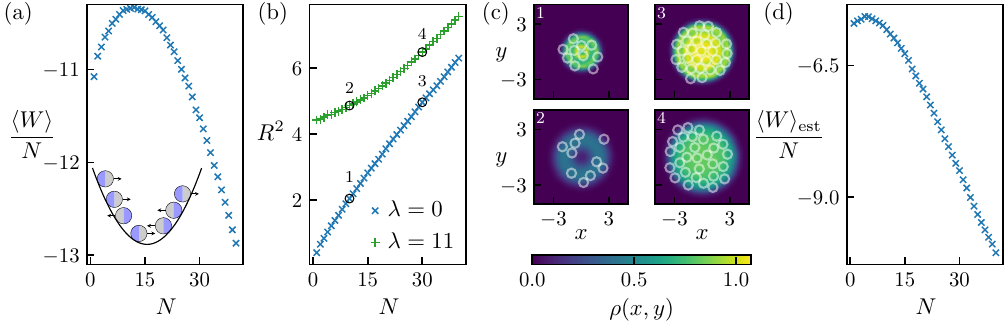} 
\caption{Many-body active engine segment. (a) Average work $\langle W \rangle$ per particle, as a function of the number of particles $N$, for neural-network protocols trained to minimize work by controlling the activity and confining potential of a set of interacting active Brownian particles. The many-body system is more efficient than the one-body system for $N>25$.  (b) Steady-state values of $R^2  \equiv N^{-1} \sum_{i=1}^N \av{r_i^2}$ as a function of $N$. Here $\kappa = 5$, and $\lambda$ is set to its minimum (blue line) and maximum (green line) values. (c) Steady-state particle density at the control parameters indicated in (b). Instantaneous particle positions for one realization of the system at steady state are shown in white. (d) A simple model of the learned protocols, \eqq{eq:work_estimate}, captures the non-monotonicity of panel (a).}
\label{fig4}
\end{figure*}

In \ff{fig3}(e) we show mean work as a function of time for the two protocols. The learned protocol requires an input of work at early times in order to extract net work at later times. This solution was identified by a genetic algorithm using an order parameter \eq{phi3} that depends only on quantities evaluated at the final time point. As a result, the protocol is not biased toward any particular functional form. By contrast, greedy reinforcement-learning algorithms, which at all times attempt to reduce the objective function, would (without special shaping of the reward function) be unlikely to find the solution shown here.

\section{Work extraction from confined, interacting active particles}\label{sec:mb}

We now consider the case of $N$ interacting active Brownian particles placed within the two-dimensional harmonic trap of \s{sec2}. Particle $i$ evolves according to the Langevin equation
\bea
\label{lang_int}
\frac{\mathrm{d}\bm{r}_i}{\mathrm{d}t} &=& \lambda \hat{\bm{e}}_i(\theta) - \kappa\bm{r}_i -\partial_{{\bm r}_i} \sum_{j \neq i} V(r_{ij})+ \sqrt{2}\bm{\xi}_r(t) \nonumber \\ 
\frac{\mathrm{d}\theta_i}{\mathrm{d}t} &=& \sqrt{2}\xi_\theta(t),
\eea
whose terms are similar to those of \eq{lang1} with the addition of the Weeks-Chandler-Andersen interaction 
\begin{equation}
V(x) = \begin{cases} 
4\epsilon\left[\left(\sigma/x\right)^{12} - \left(\sigma/x\right)^{6}\right] + \epsilon &  (x < 2^{1/6}\sigma) \\
0 &({\rm otherwise}),
\end{cases}
\end{equation}
which takes as its argument the inter-particle separation $r_{ij} \equiv |{\bm r}_j-{\bm r}_i|$. We set $\sigma$ and $\epsilon$ to 1.

We wish to learn protocols that minimize the mean work done upon reducing the trap stiffness from $\kappa_{\rm{i}} = 5$ to $\kappa_{\rm{f}} = 2$, in time $t_{\rm{f}}$, observing the bounds on the control parameter values of \s{sec2}. Here work is
\beq
    \langle W\rangle = \frac{N}{2}\int_0^{t_{\rm{f}}} \mathrm{d}t\,  \dot{\kappa} R^2,
    \label{eq:work2}
\eeq
where $R^2 \equiv N^{-1} \sum_{i=1}^N \av{r_i^2}$. The angle brackets indicate an average over dynamical trajectories. We start from a steady state at $\lambda_{\rm{i}} = 0$, but place no constraints (beyond those of the control-parameter bounds) on the value of $\lambda_{\rm{f}}$. Such a transformation could be used as part of a cycle for an active engine~\cite{fodorActiveEnginesThermodynamics2021,PhysRevE.102.010101, kumariStochasticHeatEngine2020}. 

No analytical solutions are known for this many-body system, but a protocol can be learned in exactly the same way as for the single-particle problems considered previously, using a genetic algorithm to train a neural network to minimize $\phi=\av{W}$. The latter was calculated from $10^3$ independent trajectories.

In \ff{fig4}(a) we show the result of this learning procedure for trajectory time $\tf=1$ and a number of particles between $N = 1$ and $N = 40$. In \ff{figN12} and \ff{figN40} we provide additional details of learned protocols for the cases $N=12$ and $N=40$. In all cases work can be extracted, $\av{W}<0$. However, the extracted work per particle is a non-monotonic function of $N$, attaining a minimum value for $N=12$. For this particular problem, the many-body system becomes more efficient than the one-body system for $N>25$. This finding suggests that particular cycles of many-body active engines may function more efficiently with certain particle numbers.

The learned protocols that produce the work values in \ff{fig4}(a) initially increase $\lambda$ to its maximum value. For small $N$ they initially increase $\kappa$ to its maximum value, while for large $N$ they initially increase $\kappa$ to close to it maximum value (see \ff{figN12} and \ff{figN40}). This initial increase of $\kappa$ costs work (as with the protocols in Sec.~\ref{sec3b}), but the significant increase in $R^2$ as the activity is increased allows for net work extraction upon the subsequent reduction of $\kappa$. 

As $N$ is increased from 1, the amount of work that can be extracted per particle initially goes down. This decrease results from the fact that particles repel each other, and so $R^2$ in the passive initial state is significantly larger for $N>1$ than for $N=1$; see \ff{fig4}(b). Increasing $\kappa$ (at early times) therefore costs more work per particle than for the case $N=1$. Work can still be extracted from this system, but less efficiently than for the single-body system.

For $N$ sufficiently large, however, the situation changes: $R^2$ for large $\lambda$ and $\kappa$ becomes much larger than it is for a single particle (\ff{fig4}(b,c))~\footnote{For small $N$, the value of $R^2$ for large $\lambda$ and $\kappa$ is not much larger than it is for $N=1$, because particles spread out to form a ring and can adopt a mean radial position similar to that preferred by a single particle; see \ff{fig4}(c)}. This change allows for greater work extraction per particle when $\kappa$ is decreased later on in the protocol. For $N > 25$, this effect exceeds that described in the previous paragraph, and the many-body system provides more work per particle than a one-body system.

To illustrate the origin of the non-monotonicity seen in \ff{fig4}(a), we consider a simplified protocol that instantaneously sets $\kappa$ and $\lambda$ to their maximum values, waits until the system reaches a steady state, and then sets $\kappa=\kappa\f$. The work per particle associated with this protocol is given by
\begin{align}
    \begin{split}
\label{eq:work_estimate}
\frac{\langle W \rangle_{\rm est}}{N} = \frac{1}{2}&(\kappa_{\rm{max}} - \kappa_{\rm{i}})R^2_{\lambda_{\rm{i}}, \kappa_{\rm{i}}}\\ &- \frac{1}{2}(\kappa_{\rm{f}} - \kappa_{\rm{max}})R^2_{\lambda_{\rm{max}}, \kappa_{\rm{max}}},
\end{split}
\end{align}
where $R^2_{\lambda, \kappa}$ denotes the steady-state value of $R^2$ measured at $(\lambda, \kappa)$. In \ff{fig4}(c) we show that \eqq{eq:work_estimate} is a non-monotonic function of $N$. It is not a quantitatively accurate model of the learned protocols, but captures one important feature of their behavior. 

Extending the simulation time to $\tf = 10$ allows for even greater work extraction. This improvement is achieved by a learned protocol that substantially changes the system's activity twice, from passive to active to passive again (see Fig.~\ref{fig:mb_work_SI}).

\section{Conclusions}

We have shown that evolutionary methods can train neural networks to produce efficient protocols for state-to-state transformations in simulation models of active matter. We found protocols that were more efficient than those derived recently by constrained analytical methods, and showed that neural-network methods can aid in the design of protocols that achieve extraction of work from many-body active systems. The learning scheme used here can be applied to experiment the way it is applied to simulations, suggesting a way of designing protocols for the efficient manipulation of active matter in the laboratory.\\

\begin{acknowledgments}
Work at the Molecular Foundry was supported by the Office of Science, Office of Basic Energy Sciences, of the U.S. Department of Energy under Contract No. DE-AC02-05CH11231. This research used resources of the National Energy Research Scientific Computing Center (NERSC), a U.S. Department of Energy Office of Science User Facility located at Lawrence Berkeley National Laboratory, operated under Contract No. DE-AC02-05CH11231, and the Stevin Supercomputer Infrastructure, provided by the VSC (Flemish Supercomputer Center), funded by Ghent University, FWO and the Flemish Government –- department EWI.
C.C. was supported through a Francqui Fellowship of the Belgian American Educational Foundation, and partly through the US DOE Office of Science Scientific User Facilities AI/ML project ``A digital twin for spatiotemporally resolved experiments''.
\end{acknowledgments} 

\clearpage

\title{Supplementary Information for ``Learning protocols for the fast and efficient control of active matter''}
\maketitle

\renewcommand{\theequation}{S\arabic{equation}}
\renewcommand{\thefigure}{S\arabic{figure}}
\renewcommand{\thesection}{S\arabic{section}}
\renewcommand{\thetable}{S\arabic{section}}

\setcounter{equation}{0}
\setcounter{section}{0}
\setcounter{figure}{0}
\setcounter{page}{1}

\begin{figure}[t]
\includegraphics[width=.9\columnwidth]{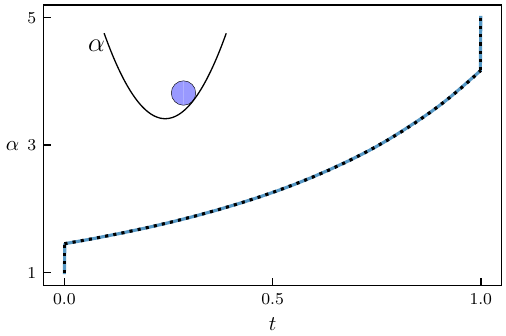} 
\caption{Benchmarking the learning algorithm: the neuroevolution procedure used in the main text was instructed to find a protocol $\alpha(t)$ of length $\tf=1$ that with least work changes from $\alpha\i = 1 \to \alpha\f = 5$ the stiffness $\alpha$ of a trap containing a passive Brownian particle. The protocol it identified is shown in blue. The black dotted line is the exact result from Ref.~\cite{schmiedl2007optimal}.}
\label{fig:seifert}
\end{figure}

\section{Benchmarking the learning procedure}
In Fig.~\ref{fig:seifert} we show the protocol $\alpha(t)$ identified by the neuroevolutionary learning procedure when instructed to minimize the work
\begin{equation}
\av{W}= \int_0^{\tf} {\rm d} t\, \dot{\alpha}(t) x(t)  
\end{equation}
for a passive Brownian particle in a harmonic trap of stiffness $\alpha(t)$, with $\alpha\i = 1$, $\alpha\f  = 5$, and $\tf = 1$.
The particle is described by (\ref{evo}) with $D_1 = 0$. The optimal protocol for this transformation is known exactly~\cite{schmiedl2007optimal}, and displays jumps at both $t=0$ and $t=\tf$. The agreement between the learned protocol and the known optimal protocol provides a benchmark for the learning algorithm when applied to the passive version of the problem studied in Section~\ref{sec:heat}, and provides confidence in the ability of the algorithm to identify rapidly-varying and discontinuous portions of protocols if necessary. For additional benchmarks see Refs.\c{whitelamDemonMachineLearning2023, whitelamHowTrainYour2023}.

\begin{figure*}
\includegraphics[width=\textwidth]{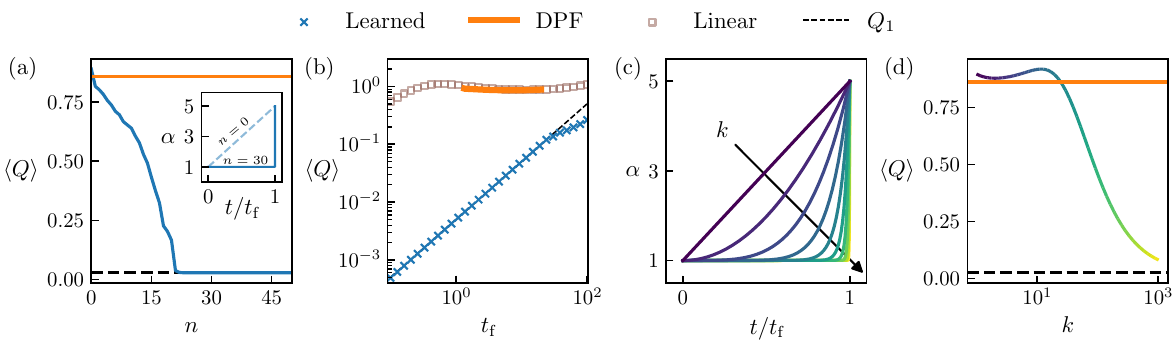} 
\caption{Supplement to \ff{fig1}. (a) Heat $\langle Q \rangle$ produced by a neural-network protocol as a function of evolutionary training time $n$, for $D_1 = 0.01$ and $\tf = 5$ (blue). The learning algorithm is instructed to minimize heat. The orange line (``DPF'') is the result extracted from Ref.~\cite{davisActiveMatterControl2023}; the dashed black line is Eq.~(\ref{q1}). Inset: the neural-network protocol before training ($n=0$) and after 30 steps of training. (b) As in Fig.~\ref{fig1}(c), but now for $D_1=0.01$. (c) Smooth protocols of the family \eq{smooth} -- with blue and green line shading indicating smaller or larger values of the exponent $k$, respectively --  and (d) associated heat values for $\tf = 5$.}
\label{fig:heat_SI}
\end{figure*}

\section{Comparison with Ref.~\cite{davisActiveMatterControl2023}}
\label{heat_si}
In Fig.~\ref{fig:heat_SI}, we compare a learned neural-network protocol that minimizes the heat $\av{Q}$ (\ref{heat}) of the system of Section~\ref{sec:heat} with the results of protocols derived in Ref.~\cite{davisActiveMatterControl2023},
in the low-activity case $D_1 = 0.01$ shown in Fig. 2 of that reference.

Fig.~\ref{fig:heat_SI}(a) shows the mean heat $\av{Q}$ produced by the neural-network protocol as a function of training time $n$. Initially the heat produced is that of the linear protocol  $\alpha(t) = \alpha \i + (\alpha \f - \alpha\i) (t/\tf)$ from which we start our search. The linear protocol produces values of heat similar to the protocols of~\cc{davisActiveMatterControl2023}, which are approximately linear. After about 20 steps of training, the neural-network protocol converges to the final-time step form, which emits an amount of heat equal to $Q_1$, given by~\eqq{q1}. 

In Fig.~\ref{fig:heat_SI}(b) we compare, for a range of values of $\tf$, the heat associated with trained neural-network protocols, the  protocols of Ref.~\cite{davisActiveMatterControl2023}, and the linear protocol. The latter two protocols produce similar values of heat, and both produce many times the heat emitted by the neural-network protocols. (For the neural-network protocols, the crossover between the forms producing heat $Q_1$ and $Q_2$ occurs here for a much larger $\tf$ than is the case for the value $D_1=2$ considered in the main text.)

The discrepancy between the protocols derived here and the protocols of Ref.~\cite{davisActiveMatterControl2023} probably results from the assumption made in that reference of slowly-varying driving, expressed by Eq. (7) of that reference, rather than from the assumption of smoothness. It is possible to construct smooth but rapidly-varying protocols that are arbitrarily close to discontinuous ones; for instance, the family of protocols 
\beq
\alpha_k(t) = \alpha\i + (\alpha\f-\alpha\i) \left(t/\tf\right)^k 
\label{smooth}
\eeq
interpolates between the linear protocol for $k = 1$ and the final-time step protocol for $k \rightarrow\infty$. We demonstrate this interpolation in Fig.~\ref{fig:heat_SI}(c) and (d) for the trajectory length $\tf=5$. The heat produced by the protocol~\eq{smooth} is not a monotonic function of $k$: a local minimum having heat values similar to those of Ref.~\cite{davisActiveMatterControl2023} occurs at $k \approx 2$. For $k \gtrsim 25$ the protocol~\eq{smooth} produces less heat than the protocols of Ref.~\cite{davisActiveMatterControl2023}, and for $k \gtrsim 10^3$ it produces heat similar to that produced by the final-time step function. 

Eq. (7) of Ref.~\cite{davisActiveMatterControl2023} represents a protocol $\alpha(t)$ as an expansion in terms of its time derivatives. Such an expansion is unable to represent rapidly-varying functions: for large $k$, near the inflection point $t \approx t_{\rm f}$ of the protocol $\alpha_k(t)$, the $n^{\rm th}$ time derivative of \eq{smooth} (for $n \leq k$) scales as $k!/(n-k)!$, and so {\em increases} in size with $n$. An expansion such as Eq. (7) of~\cc{davisActiveMatterControl2023}, which neglects derivatives above $n=2$, cannot represent such a protocol in a controlled way (the protocol $\alpha(t) \sim (t/t_{\rm f})^2$ is the strongest nonlinearity for which neglect of the third time derivative near $t=\tf$ is justified on the grounds of size). 

Our conclusion from this section and \s{sec:heat} of the main text is that methods that can represent rapidly-varying (and potentially non-analytic) protocols are needed to optimally control active-matter systems, just as they are for passive-matter systems.

\begin{figure}[b]
\includegraphics[width=\columnwidth]{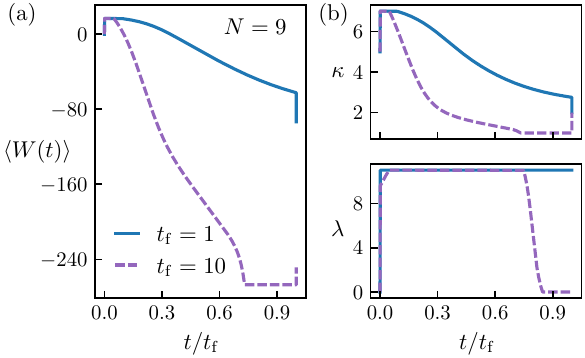} 
\caption{(a) Work as a function of time for neural-network protocols that optimize the problem of Section~\ref{sec:mb}, for $N=9$ and two values of $\tf$. (b) Protocols for $\lambda(t)$ and $\kappa(t)$ corresponding to the work in panel (a).}
\label{fig:mb_work_SI}
\end{figure}

\section{Protocols for confined interacting Active Brownian particles}

In this section we supplement the results of Section~\ref{sec:mb}. 

\begin{figure*}[t]
\includegraphics[width=0.9\textwidth]{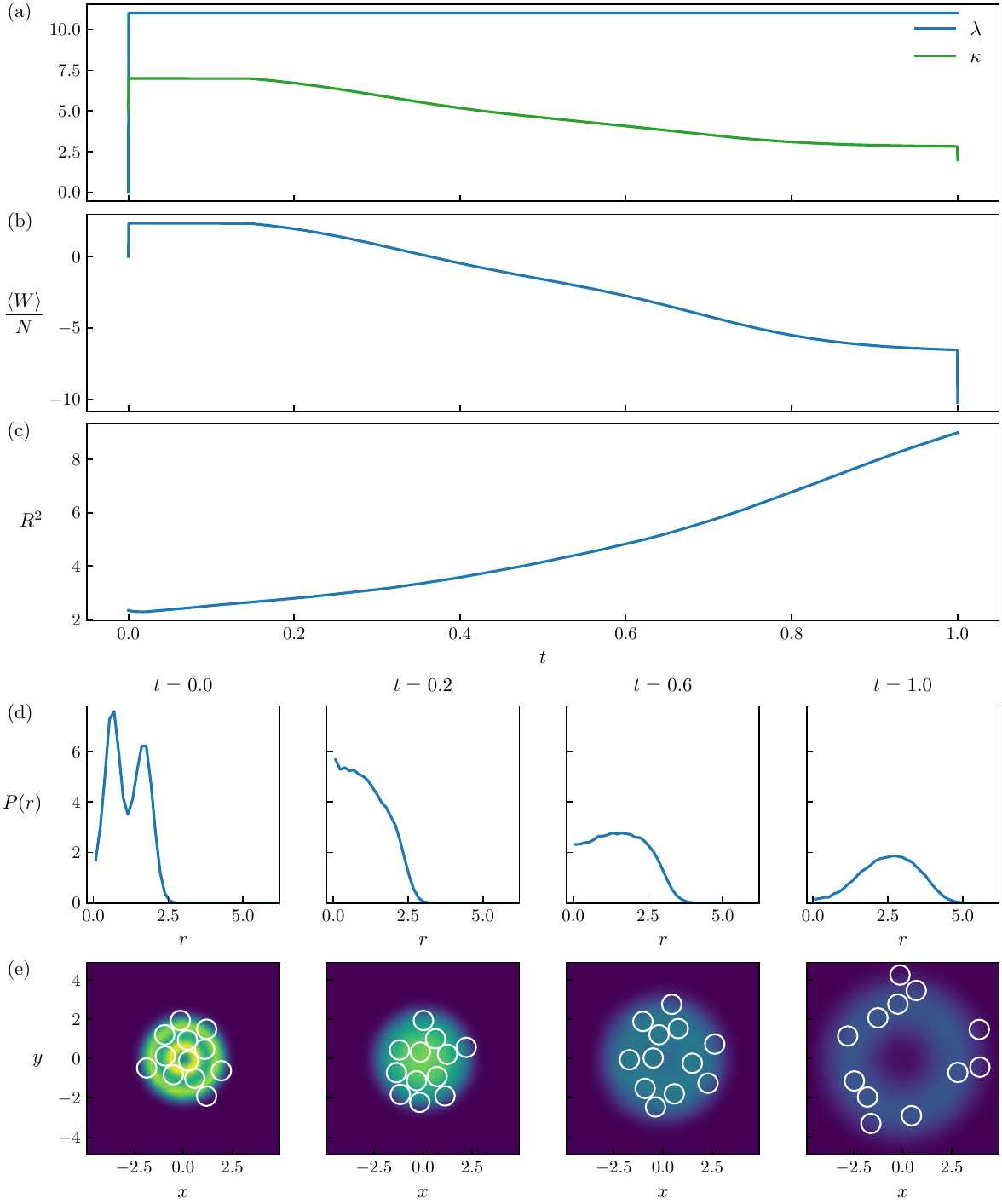} 
\caption{(a) Control parameters, (b) average work per particle, and (c) $R^2$ as a function of time for the problem of Sec.~\ref{sec:mb} and $N=12$. (d) Radial distribution function and (e) particle density averaged over $10^4$ realizations of the protocol. In panel (e) we show the particle positions for one of the trajectories. } 
\label{figN12}
\end{figure*}

\begin{figure*}[t]
\includegraphics[width=0.9\textwidth]{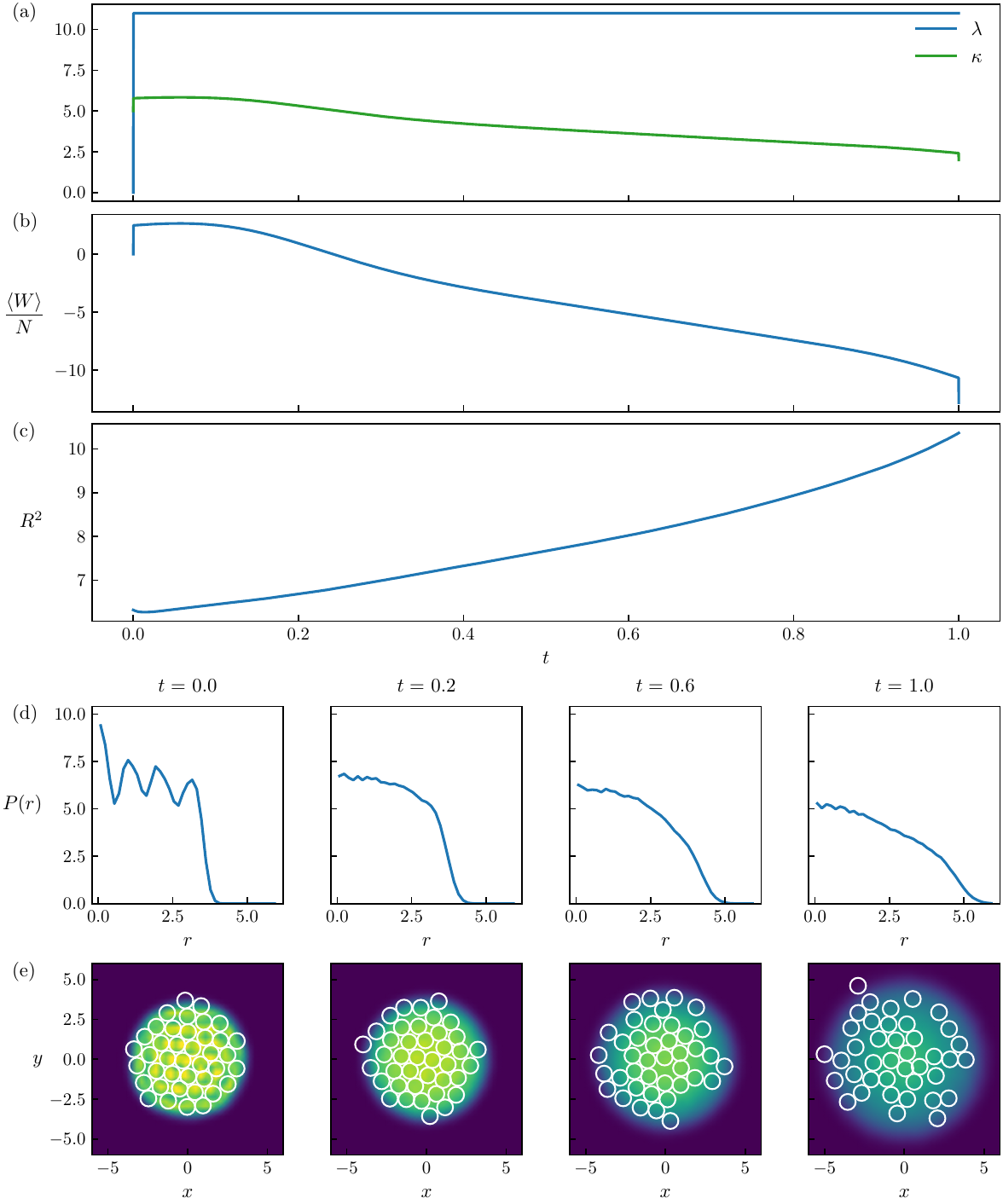} 
\caption{As Fig.~\ref{figN12}, but now for $N=40$.}
\label{figN40}
\end{figure*}

Figs.~\ref{figN12} and~\ref{figN40} provide more detail of the protocols discussed in Section~\ref{sec:mb}, for $N=12$ and $N=40$, respectively. The protocols for the control parameters $\lambda$ and $\kappa$ are shown in panel (a); panels (b) and (c) show the temporal evolution of the corresponding average work per particle and $R^2$, respectively. In panels (d) and (e) we show the radial distribution function $P(r)$ and the particle density at different times during the protocol. 

We next consider the effect of increasing the value of $\tf$ for the problem considered in that section. In panel (a) of Fig.~\ref{fig:mb_work_SI}, we show the work $\av{W}$ (\ref{eq:work}) as a function of time for a system containing $N=9$ active Brownian particles, and protocol lengths $\tf = 1$ and $\tf = 10$. The corresponding protocols for $\kappa(t)$ and $\lambda(t)$ are shown in panel (b). For $\tf = 1$, the protocol consists of increasing $\lambda$ and $\kappa$ to their maximum values instantly, and then reducing $\kappa$ while the system is in an active state, in this way extracting work.

The protocol for $\tf = 10$ is able to extract much more work than the one for $\tf = 1$, and does so by changing the state of the system twice. The protocol starts off similar to the one for $\tf = 1$, but then decreases $\kappa$ to its minimum value. Next, the system is brought back to the passive state by setting $\lambda =0$, so that the final increase of $\kappa$ to $\kappa\f$ costs less work as $R^2$ decreases.

\end{document}